\documentclass[smus]{snow2e}
\usepackage{epsfig}

\begin{document}

\begin{titlepage}
\def\today{\number\day
           \space\ifcase\month\or
             January\or February\or March\or April\or May\or June\or
             July\or August\or September\or October\or November\or December\fi
           \space\number\year}
\def\thisday{16 October 1996}
\hbox{
\vtop{
\hbox to \hsize{\hfill\large Fermilab-Conf-96/329-T}
\vspace{2.mm}
\hbox to \hsize{\hfill\large hep-ph/9610375}
\vspace{2.mm}
\hbox to \hsize{\hfill\large \thisday}}}

\vspace{1.5cm}
\begin{center}
{\LARGE \sc Distinguishing Among Models of Strong ${\rm W}_{\rm L}{\rm
W}_{\rm L}$ Scattering at the LHC
\footnote{Contributed to the proceedings of DPF/DPB Summer Study on
       New Directions for High Energy Physics, Snowmass, Colorado,
       June 25-July 12, 1996} \\[1.cm]}
{\large \sc William B. Kilgore \footnote{\tt kilgore@fnal.gov}\\[2.mm]}
{\large \it Fermi National Accelerator Laboratory\\ P.O. Box 500
\\Batavia, IL 60510, USA}

\vspace{1.5cm}
{\large \bf
ABSTRACT \\[6.mm]}
\end{center} 
{ \baselineskip=16pt \large
Using a multi-channel analysis of $W_{L}W_{L}$ scattering signals, I study the
LHC's ability to distinguish among various models of strongly
interacting electroweak symmetry breaking sectors.
}
\vfill

\end{titlepage}

\title{Distinguishing Among Models of Strong ${\bf W}_{\bf L}{\bf
W}_{\bf L}$ Scattering at the LHC}

\author{William B. Kilgore\\ {\it Fermi National Accelerator
Laboratory}\\ {\it P.O. Box 500}\\{\it Batavia, IL 60510, USA}}

\maketitle

\thispagestyle{empty}\pagestyle{empty}

\begin{abstract} 
Using a multi-channel analysis of $W_{L}W_{L}$ scattering signals, I study the
LHC's ability to distinguish among various models of strongly
interacting electroweak symmetry breaking sectors.
\end{abstract}

\section{Introduction}
The most important question in particle physics today concerns
the nature of the electroweak symmetry breaking mechanism.  One of the
most interesting and experimentally challenging possibilities is that
the electroweak symmetry is broken by some new strong interaction.  If
this is the case, there may be no light quanta (of order a few hundred
GeV or less), such as the Higgs boson, supersymmetric partners, {\it
etc\/}., associated with the symmetry breaking sector.  There will
however be an identifiable signal of the symmetry breaking sector:
strong $W_{L}W_{L}$ scattering.

The Goldstone boson equivalence theorem~\cite{ET} states
that at high energy, longitudinally polarized massive gauge bosons
``remember'' that they are the Goldstone bosons of the symmetry
breaking sector.  Accordingly, longitudinal gauge bosons in high energy
scattering amplitudes can be replaced by the corresponding Goldstone
bosons.  For weakly interacting symmetry breaking sectors, this is merely a
computational convenience.  For strongly interacting symmetry breaking
sectors, however, the equivalence theorem, coupled with the
effective-{\it W\/} approximation~\cite{ewa} becomes a powerful tool
for modeling high energy gauge boson scattering amplitudes.

Observing strong $W_{L}W_{L}$ scattering presents a very difficult
experimental challenge.  The scattering amplitudes grow with center of
mass energy, but do not become large until the mass of the
$W_{L}W_{L}$ system exceeds $\sim$1 TeV.  At the LHC, the luminosity
at such energies will be small and falling steeply, so that even
though the scattering amplitudes are large, the cross section will be
small, amounting to no more than tens of events per year.
Nevertheless, it has been shown~\cite{LHCgold,mcwk1,mcwk2} that for
all but a few pathological cases~\cite{Stealth} the LHC will be able
to establish the presence of strong $W_{L}W_{L}$ scattering, if it
exists, in at least one scattering channel.  It has also been shown
that if strong $W_{L}W_{L}$ scattering is dominated by a single
low-lying ($\sim$1 TeV) resonance, that resonance can be
identified.  The purpose of this study is to take a first look at the
difficult task of distinguishing among different models of the
symmetry breaking sector, even when there is not a single identifiable
resonance.  I will perform a multi-channel analysis on several
different models of the symmetry breaking sector, comparing the
predicted signals in each $W_{L}W_{L}$ scattering channel to those
predicted by other models.

As the basis for this study, I will use the background calculations
and signal identification cuts of Bagger {\it
et. al.\/}~\cite{LHCgold}, in which a standard set of cuts is
identified for each scattering channel, and imposed consistently on
the background processes and on a variety of models of strongly
interacting symmetry breaking sectors.
\section{$W_{L}W_{L}$ Scattering Channels}
This analysis looks at $W_{L}W_{L}$ scattering into 5 different final states:
\begin{Itemize}
\item… $ZZ \rightarrow \ell^+\ell^-\ell^+\ell^-$
\item… $ZZ \rightarrow \ell^+\ell^-\nu\overline\nu$
\item… $W^{\pm}Z \rightarrow \ell^{\pm}\nu\ell^+\ell^-$
\item… $W^+W^- \rightarrow \ell^+\nu\ell^-\overline\nu$
\item… $W^{\pm}W^{\pm} \rightarrow \ell^{\pm}\nu\ell^{\pm}\nu$
\end{Itemize}
The $ZZ \rightarrow \ell^+\ell^-\nu\overline\nu$ is included because
the small branching fraction of {\it Z\/} bosons into charged leptons
severely limits the statistical significance of the $ZZ \rightarrow
\ell^+\ell^-\ell^+\ell^-$ process.

\begin{figure}[h]
\leavevmode
\vbox{\psfig{file=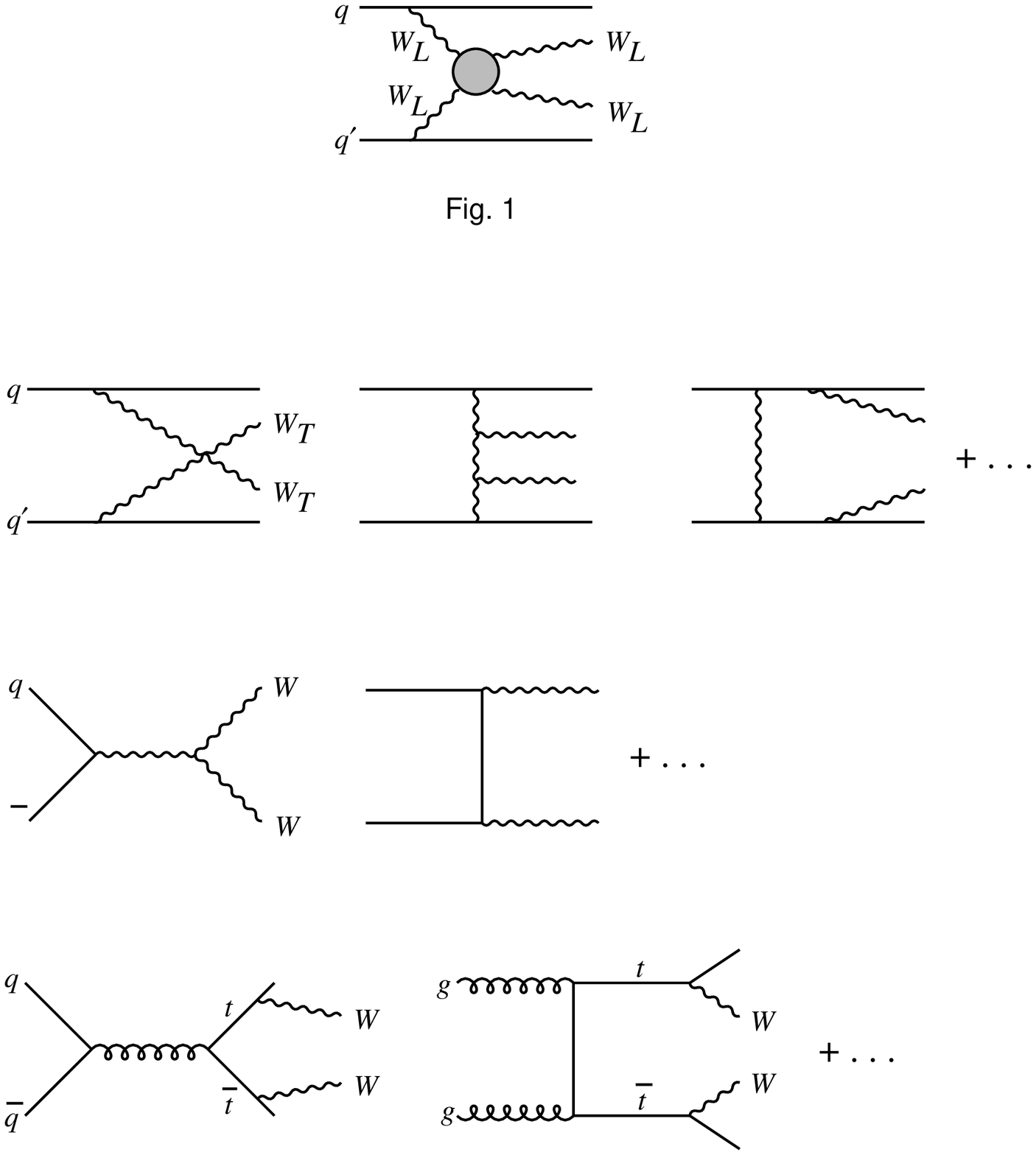,width=\hsize,clip=}}
\caption{$W_{L}W_{L}$ Fusion process}
\label{fig:wwscatt}
\end{figure}

In general, longitudinal $W_L$ pair production is dominated by the
$W_{L}W_{L}$ fusion process, in which two incoming quarks radiate longitudinal
$W_L$ bosons, which then rescatter off of one another as in
Figure~\ref{fig:wwscatt}.
Vector resonances also receive sizable contributions from
$q\overline{q}'$ annihilation into a gauge boson followed by mixing
of the gauge boson into the vector resonance state, $q\overline{q}'
\rightarrow W^* \rightarrow \rho \rightarrow W_{L}W_{L}$.

The background in this study is taken to be the standard model with a
light (100 GeV) Higgs boson.  The signal for strong $W_{L}W_{L}$
scattering is an observable excess of gauge boson pairs over the
expected rate from the standard model.  The dominant background
processes are $W_{L}W_{L}$ fusion into transverse W pairs ($qq \rightarrow
qq'W_TW_T (W_TW_L)$), $q\overline{q}'$ annihilation into W pairs plus
jets, and top quark induced backgrounds.

The strongly interacting vector boson fusion process gives the signal
events several distinctive characteristics which allow them to be
distinguished from the background.  The incoming quarks tend to emit
longitudinal gauge bosons in the forward direction which then
rescatter strongly off of one another.  The forward emission tends to
give the spectator quarks little recoil transverse momentum while the
strong scattering process, which grows stronger with increasing center
of mass energy, tends to be isotropic, throwing a large number of
events into central rapidity regions.  Thus, the signal is
characterized by high invariant mass back-to-back gauge boson pairs
accompanied by two forward jets from the spectator quarks and little
central jet activity.

This is to be contrasted with the various background processes.
$q\overline{q}'$ annihilation tends to produce transversely polarized
gauge bosons and no forward spectator jets.  When jets are produced in
association with $q\overline{q}'$ annihilation, they often appear in
central rapidities.  Top induced backgrounds tend to produce very
active events, characterized by jet activity in the vicinity of the
gauge bosons.  Perhaps the most dangerous background is the gauge
boson fusion process producing at least one transversely polarized
gauge boson since this process produces events with the same topology
as the signal.  Still, there are important differences. Interactions
involving transversely polarized gauge bosons are weak (characterized
by the weak gauge coupling) at all energies.  In order for energetic
gauge bosons to be thrown into the central region, they must typically
recoil off of the emitting quarks, rather than off of one another.
This hard recoil off of the quarks tends to throw the accompanying
jets into the central region, rather than the forward.

These signatures can be used to help formulate a set of cuts
which will enhance the signal at the expense of the background.  One
expects to find very energetic leptons in the central region of the
detector.  In addition, the leptons from one gauge boson tend to be
back-to-back with those from the other gauge boson.  In $ZZ$ modes,
the invariant mass of the $ZZ$ pair tends to be large.  In other
modes, which cannot be fully reconstructed, the transverse mass of the
gauge boson pair tends to be large.  In addition, one can veto events
with significant central jet activity, and tag for the forward
spectator jets.  The standard cuts used in Reference~\cite{LHCgold}
are summarized in Table~\ref{cutstable},
\begin{table}[t]
\centering
\caption{\label{cutstable}
Leptonic, single-jet-tagging and central-jet-vetoing
cuts for generic $W_{L}W_{L}$ fusion processes
at the LHC energy,  by final-state mode.}
\bigskip
\begin{tabular}{lcc}
\hline\hline\\[-10pt]
$ZZ($4$\ell)$ & Leptonic Cuts & Jet Cuts \\
\hline\\[-10pt]
 \ &
 $\vert  y({\ell}) \vert  < 2.5$   &
 $E_{tag} > 0.8$ TeV  \\
 \ &
 $p_T(\ell) > 40$~GeV  &
 $3.0 < \vert y_{tag} \vert < 5.0$   \\
 \ &
 $p_T(Z) > p_{cm}(Z)/2$  &
 $p_{T\ tag} > 40$~GeV   \\
 \ &
 $M({ZZ}) > 500$~GeV  & No Veto \\
\hline\\[-10pt]
$ZZ(\ell\ell\nu\nu)$ & Leptonic Cuts & Jet Cuts \\
\hline\\[-10pt]
 \ &
 $\vert  y({\ell}) \vert  < 2.5 $  &
 $E_{tag} > 0.8$~TeV   \\
 \ &
 $p_T(\ell) > 40$~GeV  &
 $3.0 < \vert y_{tag} \vert < 5.0$   \\
 \ &
 $p_T^{\rm miss} > 250$~GeV &
 $p_{T\ tag} > 40$~GeV   \\
 \ &
 $M_T(ZZ) > 500$~GeV  &
 $p_{T\ veto} > 60$~GeV \\
 \ &
 $p_T{(\ell\ell)}>M_T(ZZ)/4$ &
 $ \vert  y_{veto} \vert  < 3.0$  \\
\hline\\[-10pt]
$W^+W^-$ & Leptonic Cuts & Jet Cuts \\
\hline\\[-10pt]
 \ &
 $\vert y({\ell}) \vert < 2.0 $  &
 $E_{tag} > 0.8$~TeV  \\
 \ &
 $p_T(\ell) > 100$~GeV  &
 $3.0 < \vert y_{tag} \vert < 5.0$  \\
 \ &
 $\Delta p_T({\ell\ell}) > 440$~GeV  &
 $p_{T\ tag} > 40~$GeV  \\
 \ &
 $\cos\phi_{\ell\ell} < -0.8$  &
 $p_{T\ veto} > 30$~GeV  \\
 \ &
 $M({\ell\ell}) > 250$~GeV  &
 $ \vert  y_{veto} \vert  < 3.0$  \\
\hline\\[-10pt]
$W^\pm Z$ & Leptonic Cuts & Jet Cuts\\
\hline\\[-10pt]
 \ &
 $\vert  y({\ell}) \vert  < 2.5 $  &
 $E_{tag} > 0.8$~TeV  \\
 \ &
 $p_T(\ell) > 40$~GeV  &
 $3.0 < \vert y_{tag} \vert < 5.0$  \\
 \ &
 $ p_{T}^{\rm miss} >  50$~GeV  &
 $p_{T\ tag} > 40$~GeV   \\
 \ &
 $p_T(Z) > {1\over4} M_T(WZ) $ &
 $p_{T\ veto} > 60$~GeV  \\
 \ &
 $M_T(WZ) > 500\ {\rm GeV}$ &
 $ \vert  y_{veto} \vert  < 3.0$  \\
\hline\\[-10pt]
$W^\pm W^\pm $ & Leptonic Cuts & Jet Cuts \\
\hline\\[-10pt]
 \ &
 $\vert  y({\ell}) \vert  < 2.0 $   &
   \\
 \ &
 $p_T(\ell) > 70$~GeV   &
  $3.0 < \vert y_{tag} \vert < 5.0$  \\
 \ &
 $\Delta p_T(\ell\ell) > 200$~GeV   &
  $p_{T\ tag} > 40$~GeV   \\
 \ &
 $\cos\phi_{\ell\ell} < -0.8$  &
 $p_{T\ veto} > 60$~GeV \\
 \ &
 $M(\ell\ell) > 250$~GeV &
 $\vert y_{veto} \vert < 3.0$  \\
\hline\hline
\end{tabular}
\end{table}
\noindent
where $p_{cm}(Z)$ is the magnitude of the {\it Z\/} boson momentum in the
diboson center of mass, 
\begin{equation}
p_{cm}(Z) = {1\over2}\sqrt{M^2(ZZ) - 4M_Z^2},
\label{eq:pcm}
\end{equation}
and the transverse masses are
\begin{eqnarray}
\label{eq:transmass}
M_T^2(ZZ)&=&\left[\sqrt{M_Z^2+p_T^2(\ell\ell)} +
\sqrt{M_Z^2+|p_{T}^{\rm miss}|^2}\right]^2\nonumber\\
&& - \left[{\vec{p}}_T(\ell\ell) + {\vec{p}}_T^{\rm miss}\right]^2\nonumber\\
\\
M_T^2(WZ)&=&\left[\sqrt{M^2(\ell\ell\ell) + p_T^2(\ell\ell\ell)} +
|p_{T}^{\rm miss}|\right]^2\nonumber\\
&& - \left[{\vec{p}}_T(\ell\ell\ell) + {\vec{p}}_T^{\rm miss}\right]^2.\nonumber
\end{eqnarray}

The cuts in Table~\ref{cutstable} are chosen to maximize the
significance of each channel in the 1 TeV Higgs model.  These cuts
are not well suited for observing vector resonances in the $W^{\pm}Z$
channel.  In the Higgs model, this channel, like all others, is
dominated by the vector boson fusion process.  The cuts therefore call
for a forward jet tag.  In vector resonance models, however, more than
half of the signal in the $W^{\pm}Z$ channel comes from direct
$q\overline{q}'$ annihilation via mixing of the gauge boson and vector
resonance states.  Since these events are not accompanied by forward
spectator jets, the jet tag cuts them out of the event sample.
Reference~\cite{LHCgold} uses a special cut to enhance the $W^{\pm}Z$
signal in vector resonance models, but does not apply this cut to the
other models.
\section{Models}
\subsection{Formalism and the Lagrangian}
Models of strongly interacting symmetry breaking sectors typically fall
into one of three categories:
\begin{Itemize}
\item… Nonresonant models.
\item… Models with scalar resonances.
\item… Models with vector resonances.
\end{Itemize}
Reference~\cite{LHCgold}\ describes eight different models
of the symmetry breaking sector; three nonresonant models, three
scalar resonance models, and two vector resonance models.  The three
nonresonant models differ in the unitarization procedures imposed upon
them.  The three scalar resonance models are the standard model with a
1 TeV Higgs boson, a nonlinearly realized chiral model with a 1 TeV
scalar -- isoscalar resonance (which differs from a Higgs boson by the
strength of its coupling to the Goldstones), and an {\it O(2N)\/} symmetric
scalar interaction.  The vector resonance models incorporate vector --
isovector resonances of masses 1 TeV and 2.5 TeV in a nonlinearly
realized chiral symmetric interaction.

In this study I will use five of the models from
Reference~\cite{LHCgold}: the K-matrix unitarized nonresonant model,
the standard model, the chiral symmetric scalar resonance model and
the vector resonance models.  A single Lagrangian, transforming under
a nonlinearly realized {\it SU(2)}${}_L\otimes$ {\it SU(2)}${}_R$
chiral symmetry, can 
be written down for all of these models, with particular couplings
taking special values or set to zero as necessary.  The Goldstone
boson fields, $\pi^a$, are parameterized by the field
\begin{equation}
\xi = \exp{i{\sigma^a\pi^a\over2v}},
\label{eq:xidef}
\end{equation}
where $\sigma^a$ are the Pauli matrices and $v$ is the electroweak
vacuum expectation value.  Under chiral rotations, $\xi$ transforms as
\begin{equation}
\xi \rightarrow \xi' \equiv L\xi U^\dagger = U\xi R,
\label{eq:xitfn}
\end{equation}
where {\it L\/}, {\it R\/} and {\it U\/} are elements of {\it SU(2)\/}
and {\it U\/} is a nonlinear 
function of {\it L\/}, {\it R\/} and $\pi^a$.

With $\xi$ and its Hermitean conjugate $\xi^\dagger$, one can
construct left- and right-handed currents,
\begin{eqnarray}
\label{eq:LRcur}
J^{\mu}_L &=& \xi^\dagger\partial^{\mu}\xi \rightarrow
UJ^{\mu}_LU^\dagger + U\partial^{\mu}U^\dagger,\nonumber\\[-8pt]
\\
J^{\mu}_R &=& \xi\partial^{\mu}\xi^\dagger \rightarrow
UJ^{\mu}_RU^\dagger + U\partial^{\mu}U^\dagger.\nonumber
\end{eqnarray}
Note the inhomogeneous term $U\partial^{\mu}U^\dagger$, meaning that
these currents transform as gauge fields under the diagonal $SU(2)$.
From these chiral currents, one can form axial and vector currents,
\begin{eqnarray}
\label{eq:AVcur}
{\cal A}^{\mu} &=& J^{\mu}_L - J^{\mu}_R\rightarrow
U{\cal A}^{\mu}_LU^\dagger,\nonumber\\[-8pt]
\\
V^{\mu} &=& J^{\mu}_L + J^{\mu}_R\rightarrow
UV^{\mu}U^\dagger + 2U\partial^{\mu}U^\dagger.\nonumber
\end{eqnarray}
The axial vector current transforms homogeneously under chiral
transformation {\it U\/} but the vector current transforms
inhomogeneously.  This suggests that when we add the vector resonance
$\rho_{\mu} = \rho_{\mu}^a\sigma^a/2$, it must transform as a gauge
field under chiral transformations
\begin{equation}
\rho_{\mu} \rightarrow U\rho_{\mu}U^\dagger +
i\tilde{g}^{-1}U\partial^{\mu}U^\dagger.
\label{eq:rhotfn}
\end{equation}
Now a new vector current can be formed which transforms homogeneously
under chiral transformations,
\begin{equation}
{\cal V}^{\mu} = V^{\mu} + 2i\tilde{g}\rho^{\mu} \rightarrow U{\cal
V}^{\mu}U^\dagger.
\label{eq:Vtfn}
\end{equation}

With these pieces and a scalar -- isoscalar field $S$, we can
construct the Lagrangian,
\begin{eqnarray}
\label{eq:chilag}
{\cal L} &=& -{1\over4}v^2{\rm Tr}{\cal A}^{\mu}{\cal A}_{\mu}
- {a\over4}v^2{\rm Tr}{\cal V}^{\mu}{\cal V}_{\mu}
- {\lambda\over2}vS{\rm Tr}{\cal A}^{\mu}{\cal A}_{\mu}\nonumber\\
\\[-10pt]
&&- {1\over2}{\rm Tr}\rho_{\mu\nu}\rho^{\mu\nu}
+ {1\over2}\partial^{\mu}S\partial_{\mu}S - {1\over2}M^2_{S}S^2 + \dots,
\nonumber
\end{eqnarray}
where $\rho^a_{\mu\nu}$ is the field strength tensor of the vector
field $\rho^a_{\mu}$, and the ellipsis indicates higher derivative
terms and other terms such as couplings between the scalar resonance
and vector current which do not contribute to elastic $W_{L}W_{L}$
scattering.

In this notation, the resonances have masses and widths
\begin{eqnarray}
\label{eq:resparam}
M_S = M_S && \Gamma_S = {3\lambda^2 M_S^3\over{32\pi v^2}}\nonumber\\
\\[-10pt]
M_\rho = a\tilde{g}^2v^2 && \Gamma_\rho = {aM_\rho^3\over{192\pi v^2}}
\nonumber
\end{eqnarray}
Note that if $\lambda=1, a=0$, the scalar resonance $S$ is identical
to an ordinary Higgs boson of the standard model.  The Lagrangian in
Equation~\ref{eq:chilag} can thus parametrize a linear realization of
$SU(2)_L\otimes SU(2)_R$ even though it is written in the language of
non-linear realizations.
\subsection{Details of Particular Models}
In this analysis, I will use the results from the following models
described in Reference~\cite{LHCgold}.
\begin{Itemize}
\item The standard model with a 1.0 TeV Higgs boson ($\Gamma_S =$
0.49 TeV).  In the Lagrangian of Equation~\ref{eq:chilag}, this
corresponds to setting $M_S =$ 1.0 TeV, $\lambda =$ 1, $a =$ 0.
\item A scalar resonance with $M_S =$ 1.0 TeV, $\Gamma_S =$
0.35 TeV, corresponding to $\lambda =$ 0.84, $a =$ 0.
\item A vector resonance with $M_\rho =$ 1.0 TeV, $\Gamma_\rho =$
0.0057 TeV, corresponding to $\lambda =$ 0, $a =$ 0.208, $\tilde{g} =$
8.9.
\item A vector resonance with $M_\rho =$ 2.5 TeV, $\Gamma_\rho =$
0.52 TeV, corresponding to $\lambda =$ 0, $a =$ 1.21, $\tilde{g} =$
9.2.
\item A non-resonant model corresponding to $\lambda =$ 0, $a =$ 0.
\end{Itemize}
Note that the vector resonances considered are quite narrow.  If one
were to scale up QCD, vector resonances with masses of 1.0 and 2.5
TeV would have widths of 0.059 and 0.92 TeV respectively.  The
resonances in this study are taken to be so narrow in order to avoid
constraints on the mixing of the {\it Z\/} boson with the resonance. 
These constraints come from the effect of the vector resonance on the
spectral function of the {\it Z\/} boson.  They could be relaxed if
one were to assume, for instance, the presence of an axial vector
resonance which would have a balancing effect on the spectral
function, yet would not affect elastic $W_{L}W_{L}$
scattering~\cite{PeskTak}.
\section{Analysis}
It has been well established~\cite{LHCgold,mcwk1,mcwk2}\ that the LHC
will be able to demonstrate the existence or nonexistence of a
strongly interacting electroweak symmetry breaking sector through
direct observation of an excess of $W_{L}W_{L}$ events in at least one
scattering channel.  If such an excess is observed, one will want to
understand what sort of interaction is responsible for the excess.
Given the limited reach of the LHC into multi-TeV energies, a
realistic goal is to try to fit the observed event rates in the
various scattering channels to the predictions of various resonance
models.

To that end, I take the predicted event rates (signal plus background)
for each of the five models in turn, smear these rates by Poisson
statistics and then compare the smeared results to the expectations of
each model. By computing the mean chi-square with which the smeared
``data'' fits each model, I can determine the confidence level at
which each model can be separated from the others.

In this study, I use the event rates for a single canonical LHC year
of 100 ${\rm fb}^{\rm-1}$.  One could argue that the LHC will run for
several years and that the event rates should be multiplied by some
factor such as 3 or 5.  At present, however, I am concerned with what
can be determined in a single year of running at design luminosity and
will not speculate on the ultimate performance or lifetime of the
LHC.  The predicted event rates for the models are
shown in Table~\ref{sigtable}.
\begin{table}[h]
\centering
\caption{\label{sigtable}
Event rates per 100 ${\rm fb}^{\rm-1}$ LHC year, assuming $\protect\sqrt{s} =$
14 TeV and $m_t =$ 175 GeV.  Calculations were performed using the
MRSA parton distribution set.}
\bigskip
\begin{tabular}{c|ccccc}
\hline\hline
$\vphantom{\displaystyle{1\over2}}$\ & $ZZ(4\ell)$ & $ZZ(2\ell)$ &
$W^+W^-$ & $W^{\pm}Z$ & $W^{\pm}W^\pm$\\
\hline\\[-10pt]
Bkg. & 0.7 & 1.8 & 12 & 4.9 & 3.7\\
\hline\\[-10pt]
SM & 9 & 29 & 27 & 1.2 & 5.6\\
\hline\\[-10pt]
{\it S}\ 1.0 & 4.6 & 17 & 18 & 1.5 & 7.0\\
\hline\\[-10pt]
$\rho$\ 1.0 & 1.4 & 4.7 & 6.2 & 4.5 & 12\\
\hline\\[-10pt]
$\rho$\ 2.5 & 1.3 & 4.4 & 5.5 & 3.3 & 11\\
\hline\\[-10pt]
LET & 1.4 & 4.5 & 4.6 & 3.0 & 13\\
\hline\hline
\end{tabular}
\end{table}

\noindent Note again that the standard cuts for
the $W^{\pm}Z$ channel given in Table~\ref{cutstable} are not
optimized for the detection of vector resonances since they cut out
the half of the signal that comes from direct $q\overline{q}'$
annihilation.  Since the optimized cut is not applied to all
models, I cannot use it for a quantitative analysis.  I will however
indicate its qualitative effect on the results below.
\section{Results}
The results of the analysis are presented in Table~\ref{chisqtable}.
\begin{table*}[t]
\centering
\caption{\label{chisqtable}
Mean chi-square per degree of freedom for fitting the smeared ``data''
from each model to all of the models.  The source of the ``data'' is
indicated by the row.  The model to which it is fit is indicated by
the column.
}
\bigskip
\begin{tabular}{c|c|c|c|c|c}
\hline\hline
&Higgs & Scalar & Vector & Vector & LET-K \\
& (1.0, 0.49) & (1.0, 0.35) & (1.0, 0.0057) & (2.5, 0.52) &\\
\hline
Higgs &&&&&\\
$M_H=$ 1.0 TeV &
  $\langle\chi^2\rangle =$ 0.82 & $\langle\chi^2\rangle =$ 3.44 &
  $\langle\chi^2\rangle =$ 26.3 & $\langle\chi^2\rangle =$ 28.1 &
  $\langle\chi^2\rangle =$ 28.1\\
$\Gamma_H=$ 0.49 TeV &&&&&\\
\hline
Scalar &&&&&\\
$M_S=$ 1.0 TeV &
  $\langle\chi^2\rangle =$ 2.17 & $\langle\chi^2\rangle =$ 0.82 &
  $\langle\chi^2\rangle =$ 7.74 & $\langle\chi^2\rangle =$ 8.33 &
  $\langle\chi^2\rangle =$ 8.56\\
$\Gamma_S=$ 0.35 TeV &&&&&\\
\hline
Vector &&&&&\\
$M_\rho=$ 1.0 TeV &
  $\langle\chi^2\rangle =$ 7.72 & $\langle\chi^2\rangle =$ 3.75 &
  $\langle\chi^2\rangle =$ 0.82 & $\langle\chi^2\rangle =$ 0.93 &
  $\langle\chi^2\rangle =$ 0.95\\
$\Gamma_\rho=$ 0.0057 TeV &&&&&\\
\hline
Vector &&&&&\\
$M_\rho=$ 2.5 TeV &
  $\langle\chi^2\rangle =$ 7.51 & $\langle\chi^2\rangle =$ 3.59 &
  $\langle\chi^2\rangle =$ 0.81 & $\langle\chi^2\rangle =$ 0.82 &
  $\langle\chi^2\rangle =$ 0.86\\
$\Gamma_\rho=$ 0.52 TeV &&&&&\\
\hline
&&&&&\\
LET-K & $\langle\chi^2\rangle =$ 8.08 & $\langle\chi^2\rangle =$ 3.99 &
  $\langle\chi^2\rangle =$ 0.86 & $\langle\chi^2\rangle =$ 0.90 &
  $\langle\chi^2\rangle =$ 0.82\\
&&&&&\\
\hline\hline
\end{tabular}
\end{table*}
One can see that scalar resonance models are easily distinguished from
vector resonance and non-resonant models.  More surprising is that the
1.0 TeV Higgs boson is reasonably well separated from the narrower
1.0 TeV scalar resonance.  The reason for this is that a Higgs
theory is a renormalizable, {\it unitary\/} theory.  The couplings of
the gauge bosons to the Higgs cuts off the growth of the scattering
amplitudes in all channels and unitarizes them.  (Actually, tree level
unitarity {\it is\/} violated when the Higgs is more massive than
$\sim$800 GeV, but the theory is still renormalizable, and still
unitary when higher order corrections are considered.  The scalar
resonance model is merely a low energy effective theory and is neither
renormalizable nor unitary.)  The smaller coupling of the narrower
resonance to the gauge bosons is insufficient to unitarize the
amplitudes.

The effect of this coupling strength is easily seen from
Table~\ref{sigtable}.  The amplitudes for $W^+W^-$ and $ZZ$ production
are dominated by $s$-channel scalar exchange in the resonance region.
The smaller coupling of the narrower resonance reduces
the size of the signal in these channels.  In $W^{\pm}Z$ and
$W^{\pm}W^{\pm}$ production, {\it t\/}-channel scalar exchange reduces the
magnitude of the scattering amplitudes.  In these cases, the smaller
coupling of the resonance causes the amplitudes to be reduced less
than they would be by the Higgs, leading to larger signals.

Table~\ref{chisqtable} is somewhat misleading and overly pessimistic
in that it indicates that vector resonance models cannot be
distinguished from one another, nor from non-resonant models.  This
result is an artifact of the forward jet tag 
in the $W^{\pm}Z$ channel, which removes signal events due to
$q\overline{q}'$ annihilation.  By eliminating the jet tag and
looking in a window of transverse {\it WZ\/} mass surrounding the resonance,
the 1.0 TeV vector state can be easily identified~\cite{LHCgold}, and
the model separated from the others with a high degree of confidence.
The 2.5 TeV resonance, however, is too massive to be produced copiously,
and cannot be distinguished from non-resonant strong scattering.
Using considerably broader vector resonances,  This conclusion is
supported by References~\cite{mcwk1,mcwk2}, which have found that
vector resonances can be clearly identified in the $W^{\pm}Z$ channel
up to masses of 2.0 TeV, but that resonances above 2.5 TeV are
difficult to distinguish from non-resonant strong scattering.
\section{Discussion}
There are many ways in which this analysis can be improved.  One of
the most obvious improvements would be in the choice of cuts.  This
analysis applies the same basic set of cuts, optimized for the 1 TeV
Higgs signal, to all models.  This strategy serves the purpose for
which it was intended by setting a standard by which one can tell if
strong $W_{L}W_{L}$ scattering is occurring, but it is not well suited to the
present analysis which attempts to distinguish among models of strong
scattering.  In particular, since the $W^{\pm}Z$ signal in a Higgs
model is optimized by using forward jet tags, the cuts remove much
of the $W^{\pm}Z$ signal that occurs in a vector resonance model.  A
better analysis would optimize the cuts in each scattering channel for
each model.  One would then need to compute the performance of each
model under the other models' optimized cuts. Given a set of cuts, one
can easily compute the performance of the various models.  The
difficulty lies in performing the optimization. The detailed
background investigations that would be required are beyond the
scope of this study.

This study would also be improved by adding more models.  It would be
interesting to determine the reach for identifying vector resonances
more precisely.  It would also be interesting to look at models with
both scalar and vector resonances and study how their signal patterns
interfere with one another.

Yet another improvement on this study would be to move beyond its
reliance on gold plated purely leptonic modes.  The ATLAS and CMS
collaborations have both studied searches for 1 TeV Higgs bosons
decaying via ``silver plated'' modes, in which one gauge boson decays
leptonically while the other decays into jets, with positive
results~\cite{ATLAS,CMS}. The benefit of using the silver plated modes
is  that the hadronic branching fraction is much larger than the
leptonic branching fraction, providing a sizable increase in rate.  On
the other hand, the hadronic decay modes are much messier and
depend much more sensitively on the details of calorimetric
performance.  In addition, one cannot determine the charge of the
hadronically decaying gauge boson, obscuring the clean separation of
scattering channels.  A full investigation of the detection of silver
plated modes must await a better understanding of the actual
detectors, and will be best performed by the experimental
collaborations themselves.

\section{Conclusions}
The LHC will be able to establish the presence or absence of strong
$W_{L}W_{L}$ for most models of the strongly interacting symmetry
breaking sector.  Making use of all $W_{L}W_{L}$ scattering channels,
this analysis shows that the LHC will not only be
able to identify low lying resonances, but will also be able to
distinguish among different resonance models.  In the few models
studied here, it is apparent that resonances near 1 TeV can be
readily identified but that models with resonances above 2.5 TeV are
indistinguishable from non-resonant models.  A more definite limit on
resonance identification and ultimately on the LHC's ability to
distinguish among strong scattering models requires a more complete
analysis along the lines detailed above.
\section{Acknowledgments} I would like to thank Persis Drell and
Sekhar Chivukula for helpful comments during this analysis.  Fermilab
is operated by Universities Research Association, Inc., under contract
DE-AC02-76CH03000 with the U.S. Department of Energy. 

\vbox to 2.5cm{\vfil}
%
%

\end{document}